# Collisionless transport equations derived from a kinetic exospheric solar wind model with kappa velocity distribution functions

G. Voitcu[1], V. Pierrard[2,3], H. Lamy[2], J. Lemaire[2,3], and M. Echim[1,2]

*[1]Institute of Space Science, Magurele, Romania*

*[2]Belgian Institute for Space Aeronomy, Brussels, Belgium*

*[3]Université Catholique de Louvain, Louvain-la-Neuve, Belgium*

Corresponding author: G. Voitcu, Institute of Space Science, Romania (gabi@spacescience.ro)

**Abstract**

In this paper we discuss the collisionless transport equations, continuity, momentum and energy conservation, derived from a kinetic exospheric model of the solar wind based on a kappa velocity distribution function of the electrons. The model is stationary and is based on a non-monotonic potential energy for the protons. The present study is carried out for an exobase located at 1.5 solar radii and for two different values of the kappa index. The maximum radial distance considered is equal to one astronomical unit. The moments of the velocity distribution function computed with the kinetic exospheric model for both electrons and protons are introduced into the mass continuity equation, momentum conservation equation and energy conservation equation. The relative importance of various terms in the macroscopic transport equations for each component species are analyzed and discussed. The results obtained show that the kinetic description based on kappa velocity distribution functions satisfies rigorously the transport equations that give a macroscopic description of the solar wind plasma.

## 1. Introduction

The solar wind is the supersonic expansion of the solar corona into the interplanetary space. It is essentially a plasma of electrons and protons with a small proportion of heavier ions. The first theoretical model trying to explain the physical mechanisms that enable the acceleration of the solar wind plasma up to supersonic velocities was given by Parker (1958). Since then, many physical models have been developed to describe the properties of the solar wind measured *in-situ* on-board different spacecraft. For an extensive review of the existing kinetic and fluid solar wind models see Echim, Lemaire and Lie-Svendsen (2011).

An important characteristic of the electron velocity distribution functions (VDFs) observed *in-situ* in the solar wind is the presence of an enhanced population of suprathermal particles constituting the halo (Pierrard, Maksimovic and Lemaire, 2001a). These distributions with enhanced suprathermal tails are well represented by the kappa (or Lorentzian) functions that decrease as a power law of the velocity $v$:

$$f(v) = \frac{n}{2\pi} \left[ \frac{m}{(2\kappa - 3)k_B T} \right]^{3/2} A_\kappa \left[ 1 + \frac{mv^2}{(2\kappa - 3)k_B T} \right]^{-\kappa - 1} \tag{1}$$

where $n$ is the number density, $T$ the temperature, $A_\kappa$ a constant and $k_B$ the Boltzmann constant (see Scudder, 1992a, 1992b and Pierrard and Lazar, 2010 for a review). The value of the kappa index, $\kappa$, determines the slope of the VDF tail. The kappa function generalizes the Maxwellian case that corresponds to the limit $\kappa \rightarrow \infty$. Fits of electron distributions observed *in-situ* by Ulysses in the solar wind show that $\kappa$ ranges between 2 and 7 (Maksimovic, Pierrard and Riley, 1997).

Such distributions with suprathermal tails are general in space plasmas and suggest a universal mechanism of formation. This mechanism can be related to turbulent wave-particle interactions, as discussed by Pierrard, Lazar and Schlickeiser (2011), or to radiation pressure on atoms, ions and electrons, as described by Hasegawa, Mima and Duong-van (1985). For a discussion on the statistical description of kappa distribution see Shizgal (2007) and Livadiotis and McComas (2009).

The presence of suprathermal electrons has important consequences concerning the heating of the corona, the increase of the escape flux, the acceleration of the solar wind and the heat flux (Pierrard, 2012). More generally, the presence of suprathermal particles have important effects on the heating of all the stellar atmospheres (Scudder, 1992b; Pierrard and Lamy, 2003) and on the escape flux of particles from stellar and planetary atmospheres, from the terrestrial polar wind (Lemaire and Pierrard, 2001) to that of Saturn or Jupiter (Pierrard, 2009). The ion-exosphere model developed by Pierrard and Lemaire (1996) based on such kappa distributions has been adapted to the solar wind case by Maksimovic, Pierrard and Lemaire (1997) and then improved to cases with low exobases by Lamy *et al.* (2003).

In this paper we investigate the macroscopic transport equations in the solar wind using this kinetic exospheric model with a kappa velocity distribution function for the electrons at the exobase and a non-monotonic potential energy for the protons (Lamy *et al.*, 2003). The goal is to show and check that the moments of the non-Maxwellian velocity distribution functions given by the kinetic model fulfill the collisionless transport equations that provide a macroscopic description of the solar wind plasma.

The paper is organized as follows. In Section 2 we present an overview of the kinetic exospheric model of the solar wind developed by Lamy *et al.* (2003), while Section 3 describes the collisionless transport equations applied to the particular geometry used in our study and illustrates the results obtained. Section 4 includes our summary and conclusions.

## 2. Kinetic exospheric model of the solar wind

In this paper we use a third generation collisionless exospheric model of the solar wind described by Lamy *et al.* (2003) (see also Pierrard and Lemaire, 1996). Only protons and electrons are considered, although multi-species solutions can be treated by the kinetic approach. The model assumes that above a sharp boundary, the exobase, located at a radial distance $r_0$, the particles are fully collisionless and move freely under the action of the external forces and a given interplanetary magnetic field (IMF) distribution, **B(r)**. The external forces are the gravitation and the electrostatic force arising from a charge separation between electrons and protons due to gravitational (Pannekoek, 1922; Rosseland, 1924) and additional thermal effects. By using the Liouville's theorem together with the conservation of the energy $E$ and of the magnetic moment $\mu$ (or alternatively the speed $v$ and the pitch angle $\vartheta$), the kinetic exospheric model allows to calculate the velocity distribution function $f$ of each particle species, at any radial distance $r$ above the exobase, as a function of their distribution $f_0$ at the exobase.

Depending on the values of $v$ and $\vartheta$, the particles can be classified as either incoming, escaping, ballistic or trapped (Lemaire and Scherer, 1971). Escaping particles are those that have enough energy to escape out of the gravitational potential well of the Sun, contrary to the ballistic ones that fall back into the collision dominated barosphere. Trapped particles are bouncing back and forth between two radial distances along the same flux tube since they have not enough energy to escape at higher altitudes and their pitch angle is such that they are reflected by the mirror force at the lower altitudes, before they reach again the exobase. Incoming particles fall into the Sun from infinity and are neglected in this exospheric model.

In current ion-exospheric models the population of trapped particles is arbitrarily postulated to have a Maxwellian or kappa VDF, i.e. in detailed thermal equilibrium with those escaping from the barosphere. Of course, this is an *ad-hoc* assumption that likely overestimates the actual concentration of such trapped particles. Although more realistic density and velocity distribution functions could be envisaged in future modeling efforts, we limit the present study to the simpler case where trapped and ballistic particles are in strict thermal equilibrium and therefore have the similar energy distributions.

The model assumes a radial interplanetary magnetic field $B(r)$ varying as $r^{-2}$ (hence neglecting the Sun's rotation), although the present formulation can be extended to spiral IMF distributions like in the exospheric solar wind model developed by Pierrard *et al.* (2001).

Solar wind protons are subject to the attractive gravitational potential of the Sun $\Phi_g(r)$ and to the repulsive electrostatic potential $\Phi_E(r)$. As a consequence of the superposition of these two effects, the total potential energy of the protons reaches a maximum at some radial distance $r_{max}$ and the outward electric force exceeds the gravity force at large radial distance. Below $r_{max}$, the protons experience a global attractive potential, while all protons above $r_{max}$ can escape.

The situation is different for electrons for which the gravitational potential is negligibly small and, as a consequence, electrons are always subject to an attractive potential from the exobase until infinity. Although the gravitational force is small, it is incorporated in the general formulation of our exospheric model for self-consistency. In coronal holes the source of the fast solar wind, the exobase, is located deep in the corona because the density is much smaller there. In this case $r_0 < r_{max}$ and the solar wind protons are then located in a non-monotonic potential. At low altitudes, some protons cannot escape from the gravitational potential well of the Sun and becomes ballistic or trapped. By considering a non-monotonic potential energy for the protons, Lamy *et al.* (2003) clearly emphasized that typical characteristics of the fast solar wind can be reproduced with this model. For a more general description of the importance and impact of non-monotonic potential energy profiles on the plasma dynamics see Khazanov *et al.* (1998).

Electrons and protons VDFs both depend on the electrostatic potential $\Phi_E(r)$ which is determined self-consistently by solving iteratively the quasi-neutrality equation $n_e(r)=n_p(r)$ at each radial distance. Solving Poisson's equation is not mandatory here since the density scale height is much larger than the Debye length. The equality of fluxes of electrons and protons is also imposed in order to guarantee a zero electric current. There are three unknowns in the quasi-neutrality equation, namely the radial distance $r_{max}$ and the values of the electrostatic potential at the exobase, $V_0$, and at $r_{max}$, $V_{max}$. To find these unknowns, an iterative method is used. Details can be found in Lamy *et al.* (2003).

At the exobase, a Maxwellian VDF is assumed for protons (see Marsch *et al.*, 1982 for a detailed discussion about the protons VDF in the solar wind) and a kappa (or generalized Lorentzian) distribution is used for electrons, with suprathermal tails defined by the $\kappa$ parameter (the smaller the value of $\kappa$, the larger the suprathermal tails). There are four input parameters for the model: the radial distance of the exobase, $r_0$, the temperatures of electrons and protons at the exobase, $T_{e0}$ and $T_{p0}$, and $\kappa$. As outputs, the code provides the radial distribution of the electrostatic potential, $\Phi_E(r)$, as well as of the moments of the VDF (density, particle flux, parallel, perpendicular and total temperatures, and heat flux) of both electrons and protons.

## 3. Conservation of mass, momentum and energy of all species in the kinetic exospheric solar wind model with kappa VDF

Using the same assumptions as in the exospheric model of the solar wind developed by Lamy *et al.* (2003), i.e. steady-state conditions with spherical symmetry and radial magnetic field lines, the following equations are satisfied (Lemaire and Scherer, 1973):

- mass continuity

$$n_\alpha U_\alpha r^2 = \text{const.} \tag{2}$$

- momentum conservation

$$m_\alpha n_\alpha U_\alpha \frac{dU_\alpha}{dr} + \frac{d}{dr}(n_\alpha k_B T_{\alpha\parallel}) + \frac{2n_\alpha k_B (T_{\alpha\parallel} - T_{\alpha\perp})}{r} + m_\alpha n_\alpha \frac{d\Phi_g}{dr} + q_\alpha n_\alpha \frac{d\Phi_E}{dr} = 0 \tag{3}$$

- energy conservation

$$r^2 Q_{\alpha\parallel} + n_\alpha U_\alpha r^2 \left[ \frac{m_\alpha U_\alpha^2}{2} + \frac{k_B (3T_{\alpha\parallel} + 2T_{\alpha\perp})}{2} + m_\alpha \Phi_g + q_\alpha \Phi_E \right] = \text{const.} \tag{4}$$

where $r$ is the radial distance, $n_\alpha$ is the number density, $U_\alpha$ is the bulk velocity, $T_\alpha$ is the temperature, $Q_\alpha$ is the heat flux, $\Phi_g$ is the gravitational potential and $\Phi_E$ is the electrostatic potential; $\alpha$ is the index of the species with mass $m_\alpha$ and charge $q_\alpha$.

The five terms in the left-hand side of the momentum conservation equation (3) represent, from left to right, (i) the inertial term (further denoted T1 in Figure 5), (ii) the pressure gradient term (further denoted T2 in Figure 5), (iii) the magnetic mirror force term (further denoted T3 in Figure 5), (iv) the gravitational term (further denoted T4 in Figure 5) and (v) the electrostatic term (further denoted T5 in Figure 5).

The terms in the left-hand side of the energy conservation written in equation (4) represent, from left to right, (i) the heat flux term, (ii) the kinetic energy term, (iii) the enthalpy term, (iv) the gravitational energy term and (v) the electrostatic energy term.

The heat flux in (4) is determined in the frame of reference moving with the plasma bulk velocity. By using the energy flux, $\varepsilon_\alpha$, computed in the heliocentric frame of reference, instead of the heat flux, we obtain the following expression for the energy conservation equation:

$$\frac{\varepsilon_{\alpha\parallel}}{n_\alpha U_\alpha} + m_\alpha \Phi_g + q_\alpha \Phi_E = \text{const.} \tag{5}$$

where the first term, from left to right, is the energy flux term (denoted T1 in Figure 6), the second one is the gravitational energy term (denoted T2 in Figure 6) and the third one is the electrostatic energy term (denoted T3 in Figure 6). The energy flux in equation (5) and the heat flux in equation (4) are related by the following equation:

$$\varepsilon_{\alpha\parallel} = Q_{\alpha\parallel} + \frac{1}{2} n_\alpha U_\alpha [m_\alpha U_\alpha^2 + k_B (3T_{\alpha\parallel} + 2T_{\alpha\perp})] \tag{6}$$

The conservation of mass, momentum and energy at all radial distances between the exobase and one astronomical unit is checked for the exospheric model of the solar wind discussed in Section 2. The input parameters of the model are given in Table 1. The analysis has been carried out for an exobase located at $1.5R_s$ ($R_s$ = solar radius, $1R_s = 6.95 \cdot 10^5$ km) and for two different values of the kappa index.

The macroscopic properties of the electrons and protons determined by the moments of the velocity distribution functions assumed in the exospheric model of the solar wind are shown in Figures 1 and 2 as a function of radial distance. Two cases are illustrated: $\kappa$=2.5 (blue lines) and $\kappa$=4 (red lines).

The number density decreases asymptotically to zero and no significant differences are observed for different values of $\kappa$ (top-left panel of Figure 1). The plasma bulk velocity increases rapidly to supersonic values at radial distances less than $25R_s$, then the bulk velocity reaches an asymptotic value (top-right panel). The ion sound speed, $V_s=V_s(r)$, is marked with a dashed-line in the top-right panel of Figure 1 and it has been computed using an adiabatic index equal to $\gamma_e=\gamma_p=5/3$. The wind is accelerated to higher speeds for smaller values of $\kappa$, i.e. when there are more suprathermal electrons at the base of the corona. The electron temperature increases with altitude and reaches a maximum at a radial distance of approximately $4-5R_s$. A similar radial profile of the electron temperature has been obtained also by Zouganelis *et al.* (2004). This effect is due to the increasing of the ratio of the suprathermal over thermal particles as suggested by Dorelli and Scudder (1999) and confirmed by Pierrard and Lamy (2003). It should be mentioned that such an electron temperature maximum has been inferred by Lemaire (2012) based on radial density profiles determined from white light solar eclipse observations (see also Pierrard *et al.*, 2014). The proton temperature is monotonically decreasing with the radial distance.

The perpendicular and parallel temperatures of the protons and electrons follow the same profile of variation as the total temperature, as illustrated in Figure 2. The energy flux of the escaping protons is constantly decreasing with the radial distance; there are no significant differences for different $\kappa$. Nevertheless the energy flux of the escaping electrons is significantly larger for smaller values of $\kappa$ and two orders of magnitude larger than the energy flux of escaping protons, as in earlier kinetic and multi-fluid solar wind models. This finding confirms the key role of suprathermal electrons for the acceleration of the solar wind (Lemaire, 2010; Parker, 2010).

The gravitational and electrostatic potentials, as well as the total potential energy for protons and electrons, are shown in Figure 3. The electrostatic energy increases for smaller $\kappa$, thus an increased number of suprathermal electrons at the base of the corona means an increased acceleration potential. The link between the acceleration of the wind and the Lemaire−Scherer electric field is illustrated by the correlation between the profile of the electrostatic potential and the profile of the bulk velocity (Figures 3 and 1).

The moments of the VDF for both electrons and protons are introduced into the mass continuity equation (2), momentum conservation equation (3) and energy conservation equation (5). The derivatives in equation (3) are discretized numerically over a step-size much larger than the Debye length by using finite-differences method with a centered-difference scheme. The results obtained are shown in Figures 4, 5 and 6. The terms in the macroscopic transport equations are plotted against the radial distance from the Sun for the two different values of the kappa index, i.e. $\kappa$=2.5 and $\kappa$=4.

Figure 4 shows the variation with radial distance of the $nUr^2$ term in the mass continuity equation (2) for $\kappa$=2.5 (blue line) and $\kappa$=4 (red line). It can be seen that, for the two values of the kappa index, the $nUr^2$ term remains constant, thus the mass continuity is rigorously satisfied for both protons and electrons.

Figure 5 presents the left hand-side terms of the proton (left column) and electron (right column) momentum conservation equation (3), and also the sum of all five terms, as a function of radial distance from the Sun, for $\kappa$=2.5 (top panels) and $\kappa$=4 (bottom panels). The results obtained show that the inertial term (blue line) and the magnetic mirror force term (green line) are much smaller compared with the other ones in the case of protons, while for electrons the only significant terms are the pressure gradient term (red line) and the electrostatic term (magenta line). The sum of all five terms (black line) illustrate that the momentum conservation equation (3) is satisfied for the two different values of the kappa index, i.e. $\kappa$=2.5 and $\kappa$=4, and for both electrons and protons.

Figure 6 presents the left hand-side terms of the energy conservation equation (5) and the sum of all three terms, as a function of radial distance, for protons (left column) and electrons (right column), for $\kappa$=2.5 (top panels) and $\kappa$=4 (bottom panels). It can be noticed that the gravitational energy term (red line) is much smaller than the other two terms in the case of electrons. On the other hand, at large radial distances, the only significant term in the energy conservation equation is the energy flux term (blue line). The results obtained illustrate that the energy conservation equation (5) is well satisfied for $\kappa$=2.5 and $\kappa$=4 and for both electrons and protons.

## 4. Summary and conclusions

Exospheric models of the solar wind are exact solutions of the Vlasov equation for electrons and protons. The moments of the exospheric VDF are exact solutions of the entire hierarchy of the moment transport equations (2)−(4) without assuming a closure relationship between the moments of various orders. In this paper we describe results obtained in the collisionless approach using the exospheric model developed by Lamy *et al.* (2003). The aim of our work is to emphasize that the moments of the non-Maxwellian velocity distribution functions given by the exospheric model fulfill precisely the collisionless transport equations that provide a macroscopic description of the solar wind plasma. Note however that higher order kinetic models (e.g. Pierrard, Maksimovic and Lemaire, 1999, 2001b) include the effects of collisions by considering a Fokker-Planck collision term. It should be mentioned also that more sophisticated kinetic models have been developed by Pierrard, Lazar and Schlickeiser (2011) and Pierrard and Voitenko (2013) to include the effects of wave-particle interactions on the plasma dynamics.

The exospheric solutions show that the suprathermal electrons, described by a kappa velocity distribution function at the base of the corona, play an important role in the supersonic acceleration of the solar wind. For smaller values of $\kappa$, the wind is accelerated at higher speeds and becomes supersonic closer to the exobase. The potential energy of electrons and protons is also larger for smaller $\kappa$ and tends asymptotically to a constant that is independent of $\kappa$. This implies that at large distances the interplanetary polarization electric field vanishes and therefore does not contribute anymore to the quasi-neutrality condition.

The moments of the kappa VDF were introduced in the first three transport equations derived as moments of the Vlasov equation for electrons and protons. We obtain rigorous conservation of mass, momentum and energy at all radial distances and for all values of $\kappa$. The momentum conservation of protons is dominated by the pressure gradient term, the gravitational and electrostatic potential terms. The relative amplitude of these terms does not vary too much with $\kappa$. The conservation of momentum for electrons is mainly achieved as a balance between the pressure gradient and electrostatic terms; both terms decrease with smaller $\kappa$. The energy conservation for protons is achieved by an equilibrium between the energy flux, the gravitational and electrostatic energy terms; the first and last terms increase with decreasing $\kappa$. The results obtained here clearly illustrate for the first time the contribution of different terms in the macroscopic transport equations and their variation with the radial distance from the Sun and also with the kappa index of the electrons VDF at the exobase. On the other hand, the present study may be viewed as a reference test to crosscheck the validity of the kinetic exospheric model of Lamy *et al.* (2003) from a macroscopic perspective.

Some criticism formulated in the past suggested that exospheric models would just be "academic exercises" of little relevance to the physics of the solar wind. The fluid description is based on simplified macroscopic equations and thus it is more intuitive for solar wind modelers. Furthermore, the fluid variables are available to direct measurements. It is only recently (Lemaire, 2010; Parker, 2010) that the kinetic and hydrodynamic points of view have been admitted to be complementary and admittedly were reconciled. In this paper we show examples that clearly demonstrate how the kinetic description of the solar wind based on kappa velocity distribution functions satisfies the entire hierarchy of the moment (or transport) equations, i.e. mass continuity, conservation of momentum and conservation of energy, for both electrons and protons. In other words, the (multi)fluid equations are rigorously satisfied by the kinetic exospheric models. This is of course expected *a priori* from the basic principles of plasma physics, but an explicit demonstration in the case of kappa velocity distribution functions is certainly useful, necessary and expected by the space plasma physics community.

**Acknowledgements**

The exospheric model used in this paper is publicly available on the European Space Weather Portal (http://www.spaceweather.eu/kinetic_sw). Gabriel Voitcu and Marius Echim acknowledge support from the Romanian Ministry of Education and Research through the projects 593/2012 (OMMA) and PNII-ID-PCE-2012-4-0418 (TIMESS) and also from the European Community's Seventh Framework Programme under grant agreement no. 313038 (STORM). Marius Echim and Viviane Pierrard also acknowledge the Inter University Attraction Pole Project "CHARM" financed by the Belgian Office for Science (BELSPO). The authors thank for the support of the Wallonie-Bruxelles International (WBI) through a grant for bilateral collaboration between Romania and WBI. This study has been initiated as a student project during the "STIINTE" COSPAR Capacity Building Workshop held in Sinaia, Romania, in 2007.

**Table 1** Model parameters used in our study: $r_0$ – radial distance of the exobase in solar radii, $T_{e0}$ – electron temperature at the exobase, $T_{p0}$ – proton temperature at the exobase, $\kappa$ – kappa index for the electrons velocity distribution function at the exobase.

| $r_0$ [R$_s$] | $T_{e0}$ [K] | $T_{p0}$ [K] | $\kappa$ index | |
|---|---|---|---|---|
| 1.5 | $10^6$ | $2 \cdot 10^6$ | 2.5 | 4 |

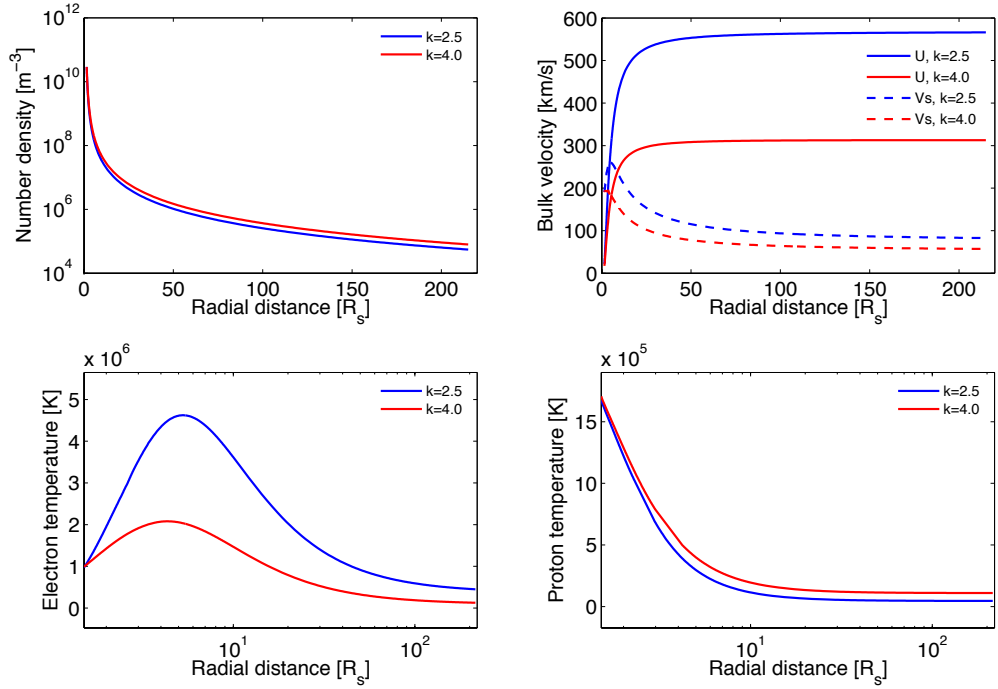

**Figure 1** Number density (top-left panel), bulk velocity (top-right panel), electron temperature (bottom-left panel) and proton temperature (bottom-right panel) as a function of radial distance for both values of the electron kappa index, i.e. $\kappa$=2.5 (blue line) and $\kappa$=4.0 (red line); the ion sound speed is shown with a dashed line in the top-right panel.

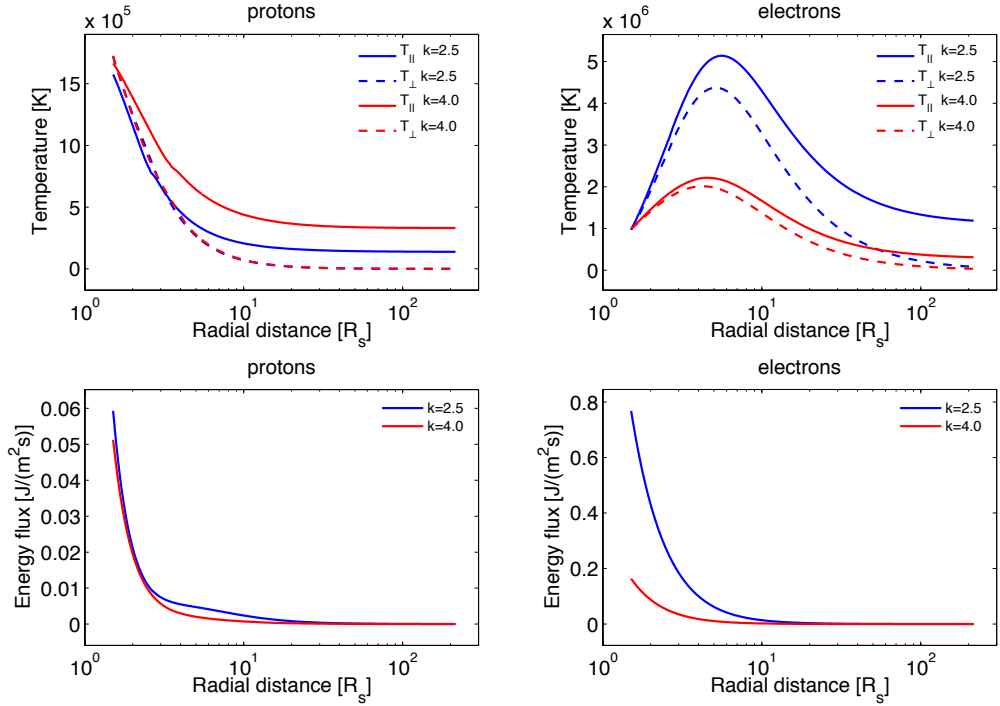

**Figure 2** Parallel (solid line) and perpendicular (dashed line) temperature (top panels) and energy flux (bottom panels) for protons (left column) and electrons (right column) as a function of radial distance for both values of the electron kappa index, i.e. $\kappa$=2.5 (blue line) and $\kappa$=4.0 (red line).

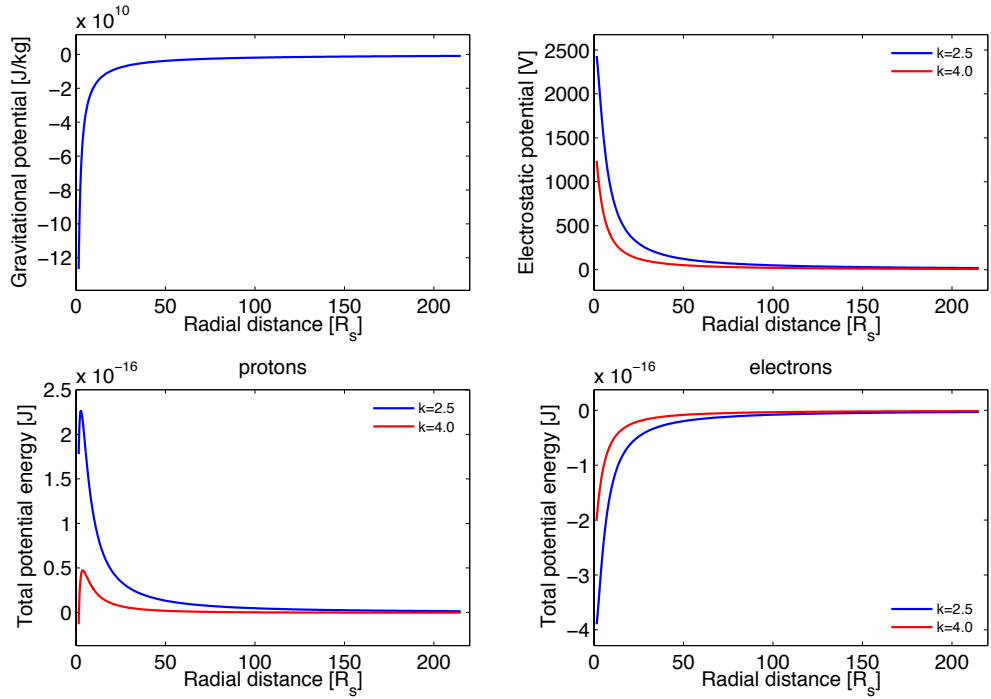

**Figure 3** Gravitational potential (top-left panel), electrostatic potential (top-right panel) and total potential energy for protons (bottom-left panel) and electrons (bottom-right panel) as a function of radial distance for both values of the electron kappa index, i.e. $\kappa$=2.5 (blue line) and $\kappa$=4.0 (red line).

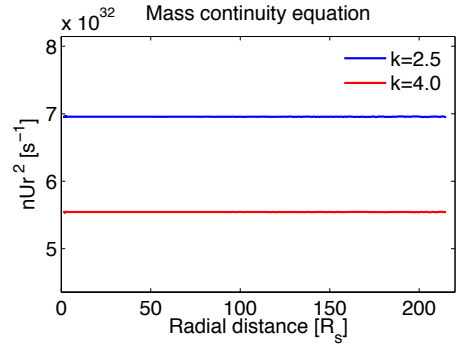

**Figure 4** Left hand-side term of the mass continuity equation (2) as a function of radial distance for both values of the electron kappa index, i.e. $\kappa$=2.5 (blue line) and $\kappa$=4.0 (red line).

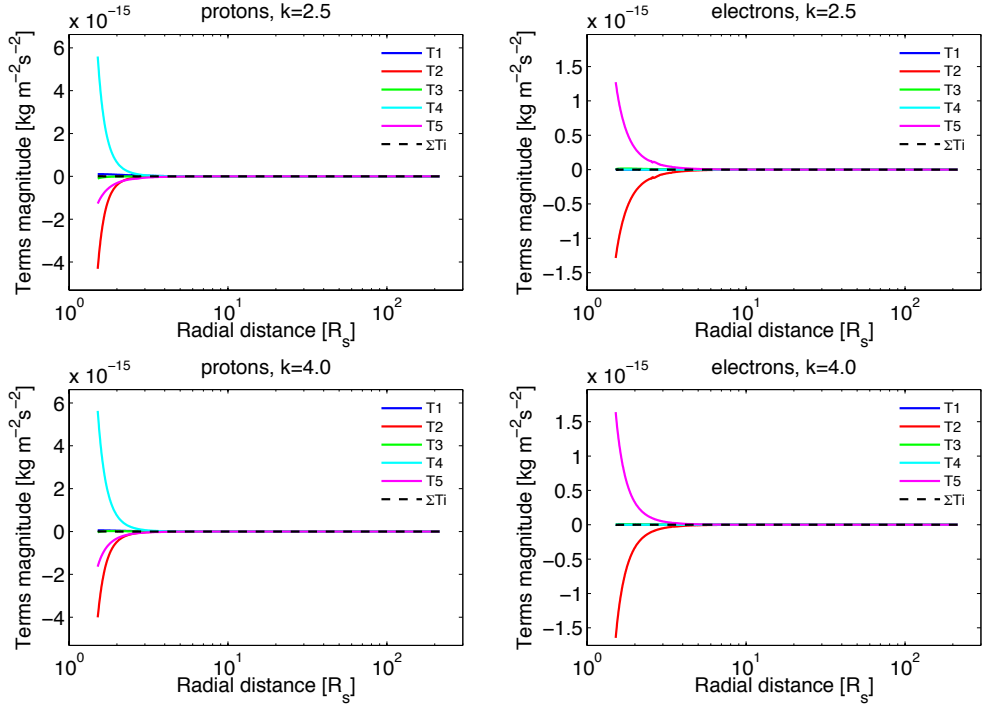

**Figure 5** Left hand-side terms of the momentum conservation equation (3) as a function of radial distance for $\kappa$=2.5 (top panels) and $\kappa$=4.0 (bottom panels), for protons (left column) and electrons (right column): T1 = inertial term (blue line), T2 = pressure gradient term (red line), T3 = magnetic mirror force term (green line), T4 = gravitational term (cyan line), T5 = electrostatic term (magenta line) and $\Sigma$Ti = sum of all terms (black line).

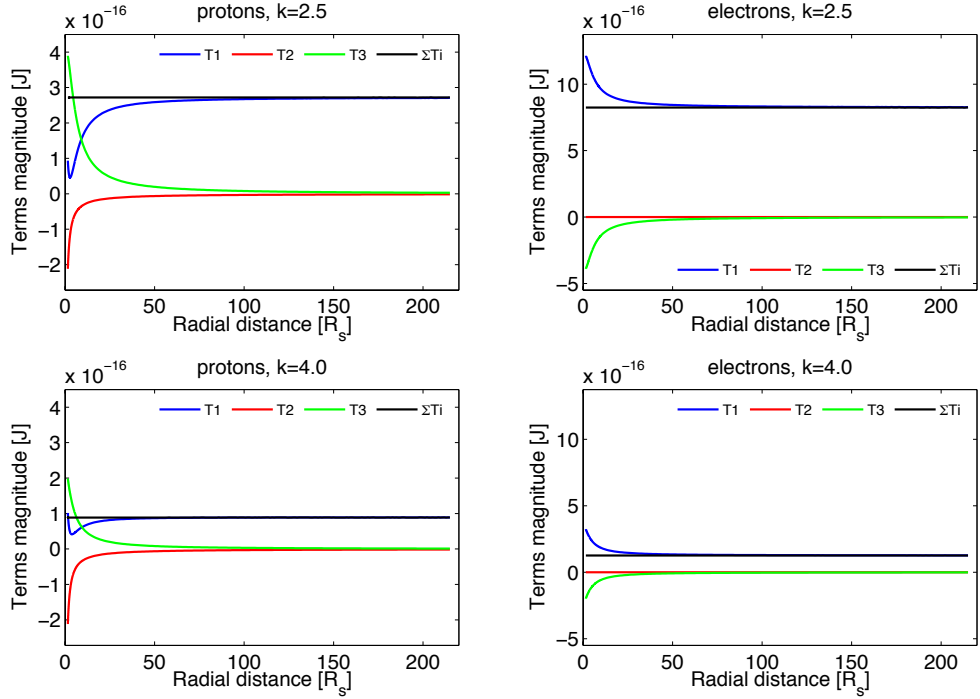

**Figure 6** Left hand-side terms of the energy conservation equation (5) as a function of radial distance for $\kappa$=2.5 (top panels) and $\kappa$=4.0 (bottom panels), for protons (left column) and electrons (right column): T1 = energy flux term (blue line), T2 = gravitational energy term (red line), T3 = electrostatic energy term (green line) and $\Sigma$Ti = sum of all terms (black line).